# METAMATERIAL COVERS OVER A SMALL APERTURE

Andrea Alù[(1, 2)], Filiberto Bilotti[(1)], Nader Engheta[(2)], and Lucio Vegni[(1)]

[(1)] Department of Applied Electronics, University of Roma Tre, 00146 Rome, Italy, alu@uniroma3.it

[(2)] Department of Electrical and Systems Engineering, University of Pennsylvania, Philadelphia, PA 19104, U.S.A., engheta@ee.upenn.edu

**ABSTRACT**

*Recently, it has been experimentally shown by several other researchers in the physics community that a properly designed periodic corrugation of a metallic opaque screen may significantly enhance the power transmission through a sub-wavelength hole in the optical frequency band. Oliner, Jackson and their co-workers explained and justified this phenomenon as the result of the excitation of the leaky waves supported by the corrugated screen. Here we suggest another setup, in which metamaterial layers with special parameters may be utilized as covers over a single sub-wavelength aperture in a perfectly electric conducting (PEC) flat screen in order to increase the wave transmission through this aperture, and we provide a detailed analytical explanation for this aperture setup that may lead to similar, more pronounced effects when the metamaterial layers are used in the entrance and the exit face of the hole in the flat PEC screen with no corrugation. Some physical insights into the function of these covers and some numerical results confirming this theory are presented and discussed. We also discuss the sensitivity of the transmission enhancement to the geometrical and electromagnetic parameters of this structure.*

## I. INTRODUCTION

Metamaterials are artificially engineered materials constructed by properly inserting many metallic or dielectric inclusions into a host material [1]-[12]. Different techniques may be employed to analyze their properties, depending on the size of the inclusions and the averaged distance among them, as compared to the wavelength of operation [1]-[3]. In particular, when both of these quantities are electrically small, so that the first term in the multipole expansion is sufficient to describe the scattering properties of every inclusion and only one Bloch mode propagates along the periodic distribution, a homogenization procedure may be applied, treating the metamaterial as a homogeneous medium with effective constitutive parameter tensors [2]. When magneto-electric coupling effects are negligible and the possible anisotropies do not affect the wave excitation under consideration, scalar effective permittivity and permeability determine the response of the bulk material [13]. Several research groups all over the world have been studying various properties of this particular class of "homogeneous" metamaterials, and in the last few years the interest has been concentrated in those media in which one or both of the constitutive parameters may attain low or negative values in certain frequency bands [3]-[12].

Independent of these developments in the area of metamaterials, in the past few years some researchers mostly in the physics community have experimentally shown that a significant enhancement of optical transmission through a periodic array of sub-wavelength holes [14] or a single sub-wavelength hole surrounded by a periodic corrugation [15] in a metallic opaque screen may be achieved by suitably choosing the corrugation periods. In both cases, they have originally attributed this phenomenon to the material surface plasmons (which is collective resonant excitations of the electron density on a surface), and in particular in the case of the single aperture they have shown how at that given period for which the transmission shows its peak, the corrugated surface supports a material polariton [15]. Oliner, Jackson and their co-workers [16]-[17] have elegantly explained this effect in terms of the leaky-wave theory, showing also how the enhancement may be optimized by a proper choice of the corrugation periods. Two important features were shown to be essential in this setup [18], [19]: a) the screen material should have a negative permittivity at the operating frequency, thus allowing presence of surface plasmons on the screen; b) the corrugation should have a certain periodicity to properly excite leaky waves.

Recently, we have presented an idea about a different setup to enhance the wave transmission through a single sub-wavelength aperture in an opaque flat perfectly electric conducting (PEC) screen, namely a case where a metamaterial slab is placed over a perfectly conducting flat screen with a small hole [20]-[21]. In the present work here, we present a detailed analysis for our proposed setup and show how a proper choice of the constitutive parameters for the metamaterial cover may lead to an analogous enhancement of transmission through the hole. In this problem, the screen may be perfectly conducting, and no corrugation or periodicity is needed. It will be shown that the leaky wave resonance at the surface of the metamaterial cover may provide similar effects as for the corrugated surface, both in reception (collecting power from an incident plane wave and directing it towards the exit side of the screen) and in transmission (increasing the directivity of wave transmission in the observer direction on the other side of the screen).

As in the case of a hole with the corrugated surrounding, this huge enhancement effect in the wave transmission may have several similar interesting applications: it may be used to spatially filter the electromagnetic radiation, effectively going even beyond the diffraction limitations (see e.g., [22]), for tunable optical filters [23], for photolithography, for near-field microscopy or to extract light from LED [15], [24]. Also the challenge of transporting optical and RF information below the diffraction limit has recently become of interest in the optics community; different solutions have been proposed [25], [26] and this effect may be another exciting way to overcome the diffraction limitation.

## II. GEOMETRY AND ORIGIN OF THE PROBLEM

The geometry of the problem is depicted in Figure 1. It is a flat perfectly electric conducting (PEC), infinitely thin metallic screen situated in a suitable Cartesian reference system. The screen is placed on the plane $y = 0$ and an electrically tiny aperture, with arbitrary cross section, is positioned at the origin. Let the region that the hole occupies on the screen be denoted by $A$. An $e^{j\omega t}$-monochromatic source is placed somewhere in the region $y < 0$, far from the hole, illuminating the structure. The screen is covered on both sides by layers of isotropic material, with complex constitutive parameters $\varepsilon$ and $\mu$ at the operating frequency $f = \omega/2\pi$, and with thicknesses $d_{in}$ and $d_{out}$ for the $y < 0$ and $y > 0$ sides, respectively. The covered screen is surrounded by free space, with permittivity $\varepsilon_o$ and permeability $\mu_0$.

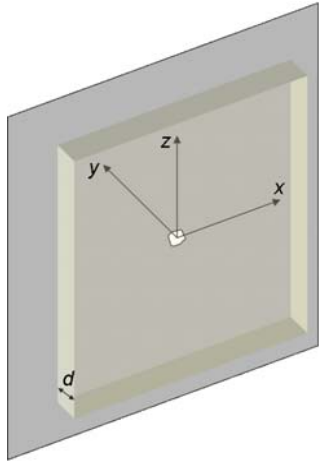

Figure 1 - Geometry of the problem: a sub-wavelength hole in a perfectly electric conducting flat screen covered by a homogeneous slab (entrance face).

First, we imagine that no cover is present on either side, i.e., $d_{in} = d_{out} = 0$ or $\varepsilon = \varepsilon_0$ and $\mu = \mu_0$. Provided that the linear dimension of the hole is sufficiently small (as compared with the wavelength), most part of the power coming from the source and impinging on the screen is obviously reflected back, due to its infinite conductivity, and only a small fraction of the incident power tunnels through the hole and is radiated in the region $y > 0$. Therefore, an observer located at $\overline{P} \equiv (x = 0, y = \overline{y}, z = 0)$, i.e., at the broadside on the other side of the screen in the radiation region would measure a very low received power. This power, and the radiation pattern from the hole, may be evaluated with very good approximation by the theory presented by Bethe [27] and reported with some extensions by Jackson [28].

According to Bethe's theory, we should first neglect the presence of the hole (i.e., we should "close" the hole) and then evaluate the field induced by the excitation in the region A, since its small dimensions should not sensibly affect the field on the entrance face. Then, we may calculate the equivalent magnetic and electric sources to be put on the other side of the screen, again in the region A, in order to solve the radiation problem on the exit face. As shown by Bethe [27], these sources are directly proportional to the amplitude of the field in region A when the hole is closed, and for an observer far from the hole on the exit side they are represented by an electric and a magnetic dipole as follows:

$$\begin{aligned} \mathbf{p} &= \alpha_1 \mathbf{E}_0 \\ \mathbf{m} &= \underline{\alpha}_2 \cdot \mathbf{H}_0 \end{aligned} \quad (1)$$

where the electric polarizability $\alpha_1$ and the magnetic polarizability tensor $\underline{\alpha}_2$ depend on the shape of the hole, and owing to the small dimensions of A, they may be evaluated using the quasi-static approximations [28]. In particular, for symmetrically shaped holes $\underline{\alpha}_2$ reduces to a scalar. The fields $\mathbf{E}_0$ and $\mathbf{H}_0$ represent, respectively, the uniform electric and magnetic fields present in region A in the absence of the hole, which implies the normal component of the total electric field and the tangential component of the total magnetic field. The polarizability factors in (1) reduce drastically with the averaged radius a of the hole and the total transmission for narrow holes is proportional to $(a/\lambda_0)^4$ [27], which represents the main limit in near-field optics.

In the following, we look for a way to enhance the wave transmission through the hole without increasing its dimensions. One way is the one experimentally shown in [15], [19], [22] and theoretically explained by Oliner, Jackson and their co-workers [16]-[17]. They have shown how a system supporting one or more leaky waves at the frequency of interest may be created by corrugating the surface of a negative-permittivity opaque screen (which is the case when metals such as silver and gold are used in the visible or infra-red (IR) regimes). The leaky waves may couple with the excitation, increasing the total transmission through the narrow aperture. In their setup, as explained in [16]-[17], the process works in two steps: the corrugation at the entrance face would help "collecting" power into the hole, by coupling the excitation with the supported leaky waves; the excitation of the leaky waves due to the corrugation on the exit face of their setup would essentially reshape the beam concentrating the power in narrow beams [16]-[17].

In our proposed geometry, we show how it is possible to obtain similar, and even more pronounced, effects employing artificial metamaterials, with no need for periodic corrugations and at frequencies (e.g., microwaves) for which the metal screen behaves as a perfect electric conductor.

## III. HEURISTIC INTERPRETATION

We now put back the cover on the screen, as in Figure 1, and for the moment we concentrate on the exit face. If the cover is a standard lossless dielectric, with $\varepsilon > \varepsilon_0$ and $\mu = \mu_0$, the radiation from the hole would be very low and with a broad beam. A sketch in terms of ray theory is given in Figure 2a. From the figure, it is obvious how a part of the power is trapped in the dielectric layer as the surface waves supported by the structure, corresponding to the rays that have total internal reflection at the dielectric-air interface. This setup, of course, would not help reshaping the pattern on the exit face in order to produce a highly directive radiation pointing towards the observer located at broadside: the radiation pattern, depicted schematically in the Fig. 2a, is broad. By reciprocity, it can be shown that such a cover on the entrance face would not help enhancing the power collected into the hole and the total power received by the observer on the exit side of the screen remains very low.

The situation changes when we employ as a cover a lossless metamaterial with a very low value of $|k| = \omega|\sqrt{\mu\varepsilon}| \ll k_o = \omega\sqrt{\mu_o\varepsilon_o}$. Such a material may be constructed at microwave frequencies, for instance, by filling a host dielectric with thin conducting rods with a proper geometry, (see [3], [4], [8]), thus lowering its effective permittivity. Provided that the lattice has a three-dimensional structure, its expected behavior is quasi-isotropic, so that after a homogenization process we may consider the material as a composite medium with effective permittivity $\varepsilon$ and $|\varepsilon| \ll \varepsilon_0$. Also other techniques, similar to those applied to build electromagnetic band-gap structures have been proposed to synthesize plasma-like artificial materials at microwave frequencies [11]. Analogously, materials with low $\mu$ may be similarly constructed, for instance, by embedding ring inclusions in the host substrate, showing analogous effects in the electromagnetic wave interaction [5], [6].

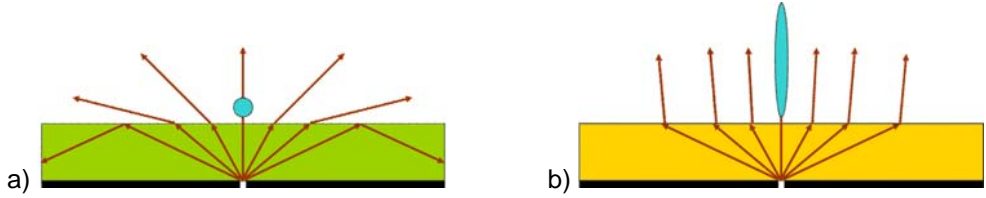

Figure 2 - Heuristic ray-theory interpretation of the behavior of the structure of Figure 1 with a slab covering the exit face: a) When a standard lossless dielectric with $\varepsilon > \varepsilon_0$ is employed (broad radiation pattern), b) When a metamaterial slab with $|k| \ll k_0 = \omega\sqrt{\varepsilon_0\mu_0}$ is employed (highly directive pattern).

As evident from Figure 2b, in this case we expect to have a much more directive radiation pattern in the broadside direction, similar to what has been studied in [9]-[12] for antenna applications. The wave-number in the medium, in fact, is much less than the one in the vacuum half-space and at the exit face of the cover slab we expect to have a small phase difference between field values at distant points on the slab surface. Therefore, due to the metamaterial slab with $|k| \ll k_0 = \omega\sqrt{\varepsilon_0\mu_0}$, the pattern becomes more directive, and an observer placed at broadside would collect much more power than the case with no cover. This effect is similar to the highly directive slabs with low permittivity presented along the years in the literature [9]-[12]. In the mathematical limit of $k \to 0$, at the exit face of the slab we expect to have an equi-phase surface and the radiation pattern would mathematically tend to a delta function in the broadside direction.

The reciprocity assures that the same setup should also be effective at the entrance face of the screen in "collecting" power into the hole from a plane wave impinging at broadside, and the whole system would heuristically work as illustrated in Figure 3.

From this heuristic explanation, it is clear that the enhancement effect is essentially related to the leaky-wave radiation properties of a slab with low $|k|$: at the entrance face the cover acts as a receiving antenna, collecting power into the hole, whereas at the exit face it operates as a transmitting antenna, directing the tunneled power towards the observer, analogous to the explanation given by Oliner, Jackson and their co-workers for the corrugated screen [16]-[17]. The hole in this design acts as a coupling aperture for the two "antennas". In the next section we will show a detailed analysis to support this heuristic explanation and we will find the conditions under which an optimum enhancement may be produced.

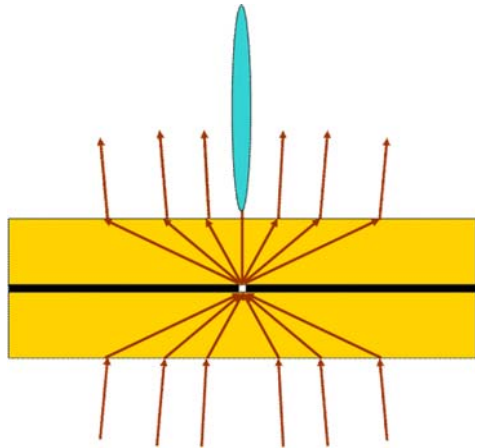

Figure 3 - Heuristic ray-theory interpretation of the dramatic enhancement using metamaterial cover slabs with low $|k|$ on both sides of the hole.

## IV. DESIGN FOR THE METAMATERIAL COVERS

### a) METAMATERIALL COVER IN THE ENTRANCE FACE

From Eq. (1), we note that in order to analyze the transmission through the tiny hole we should evaluate the tangential component of magnetic field $\mathbf{H}_0$ and the normal component of electric field $\mathbf{E}_0$ induced by our excitation on the entrance face of the screen in the region $A$, in the absence of the hole. The cover slab is positioned in the region $-d_{in} \leq y \leq 0$ and is assumed to be infinitely extent in the $x$ and $z$ directions. Provided that our excitation is far from the screen, its spectrum impinging on the grounded slab is composed only of propagating waves along the direction $y$, i.e., it may be expressed as an integral sum of $TE^y$ and $TM^y$ $e^{-jk_x x - jk_y y - jk_z z}$ plane waves with $0 < \sqrt{k_x^2 + k_z^2} = k_t < k_0 = \omega\sqrt{\varepsilon_0 \mu_0}$. Each one would induce on the surface of the screen (and in particular in the region $A$) a tangential magnetic field with complex amplitude given by:

$$H_0^{TE}(k_t) = \frac{2\omega^{-1} c_0^{TE} k_{y0} k_y e^{jk_{y0}d - jk_x x - jk_z z}}{\mu_0 k_y \cos(k_y d_{in}) + j\mu k_{y0} \sin(k_y d_{in})}$$

$$H_0^{TM}(k_t) = \frac{2\eta_0^{-1} c_0^{TM} k_{y0} \varepsilon e^{jk_{y0}d - jk_x x - jk_z z}}{\varepsilon k_{y0} \cos(k_y d_{in}) + j\varepsilon_0 k_y \sin(k_y d_{in})}$$

(2)

where $k_{y0} = \sqrt{k_0^2 - k_t^2}$, $k_y = \sqrt{k^2 - k_t^2}$, $k = \omega\sqrt{\varepsilon\mu}$, $\eta_0 = \sqrt{\mu_0/\varepsilon_0}$. The coefficients $c_0^{TE/TM}$ are the amplitudes of the electric field (expressed in V/m) of every incident plane wave. These amplitudes depend on $k_t$ and represent the spectral contributions of the excitation, obtained after a spatial Fourier transform. Notice that for the impinging propagating waves, $k_t$ is real and $k_t < k_0$, therefore $k_{y0}$ is real as well, and heretoafter we consider its positive root to satisfy the radiation condition. For what concerns the choice of sign for $k$ when negative parameters are employed, the reader is referred to [7]. Its value, moreover, may be complex when the loss in the cover materials is considered. The choice of sign for the square root in $k_y$ does not influence the field expressions and the following analysis. The analogous expressions for the complex amplitudes of the normal electric field on the screen are given by:

$$E_0^{TE}(k_t) = 0$$

$$E_0^{TM}(k_t) = \frac{2\omega^{-1} \eta_0^{-1} c_0^{TM} k_{y0} k_t e^{jk_{y0}d - jk_x x - jk_z z}}{\varepsilon k_{y0} \cos(k_y d_{in}) + j\varepsilon_0 k_y \sin(k_y d_{in})} .$$

(3)

From Eq. (1), it follows that an enhanced transmission through the hole is related to a high value of $E_0$ and $H_0$, since they are proportional to the amplitudes of the equivalent dipoles radiating on the exit side of the screen. This is consistent with the high electromagnetic fields experimentally measured by others in proximity of the surface of the metal screen when the peaks in the transmission are verified in the setups of [15], [18], [24].

As expected, the denominators in (2) and (3) represent the dispersion relations for the natural modes of this setup and, when they approach zero, the field in the slab, and in region A in particular, tends to an infinitely large value, leading to the possibility of a strong wave transmission through the hole. In our case, since we consider only impinging traveling (i.e. not evanescent) plane waves with $k_t < k_0$, these dispersion relations do not have real solutions for $k_t$, but only complex ones (i.e. $k_t = k_{tr} - jk_{ti}$), which correspond to improper leaky modes when $|k_{tr}| < k_0$ and $k_{ti} > 0$ [28]. These natural modes radiate energy at angles $\theta = \arcsin(\pm|k_{tr}|/k_0)$ from the normal to the slab, and this energy leakage explains their amplitude decay along the slab, as manifested by the quantity $k_{ti}$. Interestingly, the imaginary part, $k_{ti}$, of the complex solutions of the dispersion relations may become sufficiently small for a judicious choice of the slab parameters (permittivity, permeability, thickness), so that the corresponding natural mode couples very well with the impinging incident wave having a transverse wave number $\bar{k}_t \cong k_{tr}$. The overall effect would be to induce on the surface of the screen a strong enhancement of the field related to that component and consequently a strong power transmission enhancement through the tiny hole.

By inspection and after some algebraic manipulations of the denominators in (2) and (3), the following necessary and sufficient conditions on the entrance slab for supporting a natural mode with this required behavior are obtained:

$$|\mu| \ll \mu_0, \quad d_{in} = \frac{(2N-1)\pi}{2\sqrt{k^2 - \bar{k}_t^2}}, \qquad (4)$$

$$|\varepsilon| \ll \varepsilon_0, \quad d_{in} = \frac{N\pi}{\sqrt{k^2 - \bar{k}_t^2}}, \qquad (5)$$

where N is a positive integer and the formulas are valid in the limit of lossless materials for both polarizations. In the case of poor conductors, they still remain valid, but of course the real part of k should be considered.

These two design formulas require low-permittivity or low-permeability materials, which usually have $|k| \ll k_0$. Since $\bar{k}_t < |k|$ for the formulas to be acceptable, a good coupling may be obtained in this case only with plane waves incident close to broadside, as anticipated heuristically by Figure 2b. However, provided that the two constitutive parameters may be independently chosen, we may design for instance a sufficiently high $|\varepsilon|$ for a metamaterial with a very low $|\mu|$ satisfying (4), in order to have a value of $|k|$ that allows a wider spectrum of "coupling" angles. The same may be said for relation (5).

This particular type of leaky waves, which show a low damping factor and concentrate the field in the material cover, are strictly connected to the material polaritons (i.e., natural modes) supported by the grounded slab [30]-[32]. Their dispersion relation, which may be evaluated with a technique similar to the one proposed in [30]-[32], are given by:

$$\begin{aligned} \text{TE:} & \quad \frac{k_y}{\mu} \cot(k_y d_{in}) = -\frac{k_{y0}}{\mu_0} \tan(k_{y0} d_{in}) \\ \text{TM:} & \quad \frac{k_y}{\varepsilon} \tan(k_y d_{in}) = -\frac{k_{y0}}{\varepsilon_0} \cot(k_{y0} d_{in}) \end{aligned} \quad (6)$$

and their solutions for $k_t$, all real, consistently tend to the real part of the poles of (2) and (3) in the limit of (4) or (5), which are the cases we are interested in.

Table I reports the value of the field enhancement factors $R_E = \frac{|E_0|_{\text{with cover}}}{|E_0|_{\text{no cover}}}$ and $R_H = \frac{|H_0|_{\text{with cover}}}{|H_0|_{\text{no cover}}}$ (in agreement with the notation adopted in [16]-[17]) for the fields on the hole in the two cases of covers following condition (4) or (5) (with respect to the no-cover case) for the two polarizations (and also in particular for the normal incidence). Notice that for the TE polarization and for normal incidence no electric field is induced on the screen and therefore the enhancement factor cannot be evaluated.

Table I - Enhancement factors $R_E$ and $R_H$ for slabs satisfying conditions (4) or (5). $\eta = \sqrt{\mu/\varepsilon}$ is the intrinsic impedance of the material cover.

|  | Condition (4) | | Condition (5) | |
| --- | --- | --- | --- | --- |
|  | $R_H$ | $R_E$ | $R_H$ | $R_E$ |
| TE | $\lvert(\mu_0 k_y)/(\mu k_{y0})\rvert$ | Not Applicable | 1 | Not Applicable |
| TM | $\lvert(\varepsilon k_{y0})/(\varepsilon_0 k_y)\rvert$ | $\lvert(\varepsilon k_{y0})/(\varepsilon_0 k_y)\rvert$ | 1 | 1 |
| Normal incidence | $\eta_0/\lvert\eta\rvert$ | Not Applicable | 1 | Not Applicable |

As evident from the table, not all the possibilities work equally well for our purpose. This is due to the fact that even if the material polaritons supported by slabs satisfying (4) or (5) all concentrate the field inside the metamaterial slab, they do not necessarily have the maximum field amplitude *on* the screen (with the hole closed). In particular, the polaritons supported by slabs satisfying condition (5) usually show a very low field on the screen and their maximum value for $R_E$ and $R_H$ is equal to unity (i.e., only at the polariton resonance the field on the screen equals the value of the field without cover). The polaritons supported by this type of slabs, in fact, have their field peaks not *on* the screen, but $\lambda/4$ apart from it, at the centre of the cover slab. It follows that covers designed following condition (5) are not of interest to enhance the wave transmission through the hole. In the following, therefore, we will focus our attention on slabs designed through condition (4). (If instead of the PEC screen we had considered a perfectly magnetic conducting (PMC) screen, by duality the corresponding condition (5) would have become more suitable.)

The first design we consider here corresponds to the case of an expected radiation impinging at broadside ($k_t = 0$). The cover slab is consequently designed from equation (4) so that its leaky dominant mode radiates at broadside, i.e., with $\bar{k}_t = 0$ in (4). For the parameters used ($\varepsilon = \varepsilon_0$, $\mu = 10^{-3}\mu_0$, $d_{in} = \pi/2k$), the complex root of the TE dispersion relation (denominator of the first of (2)) is $k_t^{TE} = (4.464 - j4.51)\cdot 10^{-3} k_0$ and the corresponding TE material polariton has $k_t^{TE} = 5.228\cdot 10^{-3} k_0$ (solution of the first of (6)). In Figure 4, the factor $R_H$, which is the only one of interest in this case, is plotted versus the variation of $k_t$ of an incident TE plane wave (i.e. the variation of its incidence angle). As expected, the peak for the field enhancement is located at broadside ($k_t \simeq 0$) and it rapidly decreases as the plane wave changes its angle of incidence. In the plot only the values of $k_t < k \ll k_0$ have been reported. Notice that, consistently with the analysis developed in [16]-[17], the leaky mode cannot radiate exactly at broadside, since the condition $k_{tr} = 0$ corresponds to its cut-off. After an inspection of the TE dispersion relation, it is evident that a leaky mode with $k_{tr} = 0$ may be supported only if $\varepsilon$ or $\mu$ are identically zero, which is consistent with the heuristic interpretation given in the previous section. (As an aside it is worth noting that the possibility of having $\varepsilon$ and/or $\mu$ identically zero at a given frequency has been addressed in recent papers, e.g. [33], and some electromagnetic properties of these peculiar media have been already investigated). As reported in Table I, for the case at hand, the value of the peak for $R_H$ is $\eta_0 / |\eta| \cong 33$.

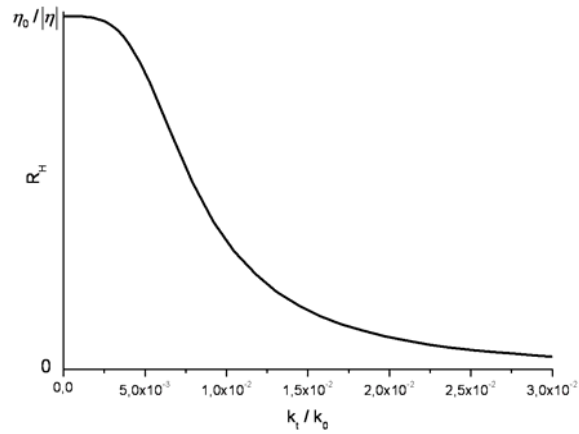

Figure 4 - Variation of the enhancement factor $R_H$ versus $k_t$ of a TE incident plane wave. The cover parameters are $\varepsilon = \varepsilon_0$, $\mu = 10^{-3} \mu_0$, $d_{in} = \pi / 2k$, which correspond to transverse wave-numbers $k_t^{TE} = (4.464 - j4.51) \cdot 10^{-3} k_0$ and $k_t^{TE} = 5.228 \cdot 10^{-3} k_0$, respectively for the leaky wave and the material polariton supported by the grounded slab.

It should be noted that the required thickness of the cover slab may be quite large (in the case of Figure 4, $d \simeq 8\lambda_0$, where $\lambda_0 = 2\pi / k_0$ is the wavelength in free space). A way to decrease this value is to play with the other constitutive parameter (in this case $\varepsilon$). Provided that it is technologically possible to increase the value of $\varepsilon$ while keeping the value of $\mu$ small, an increase in $\varepsilon$ would have the advantages of: a) reducing the electrical thickness of the cover; b) increasing the range of angles for which a good coupling may be designed; and c) increasing the enhancement factor $\eta_0 / |\eta|$. Moreover, it is interesting to add here that we have recently proposed another alternative way of reducing such cover thickness by using metamaterial bilayer covers [21]. In a future work, we will present a detailed analysis of such a case.

Referring to the same setup, in Figure 5 the spatial variation of the electric and magnetic field amplitudes versus y is sketched for an incident TE plane wave whose electric field is given by $\mathbf{E}(x,y) = \hat{\mathbf{z}} e^{-jk_t x} e^{j\sqrt{k_0^2 - k_t^2} y}$, with a transverse wave-number equal to the TE material polariton one $k_t = 5.228 \cdot 10^{-3} k_0$. The complete coupling between the plane wave and the cover polariton assures that the reflected field from the slab (and possibly the corresponding enhancement on the screen) has its maximum and that the total field is entirely dominated by the material polariton distribution, as evident from Figure 5 where the field complex amplitudes have only real or imaginary parts.

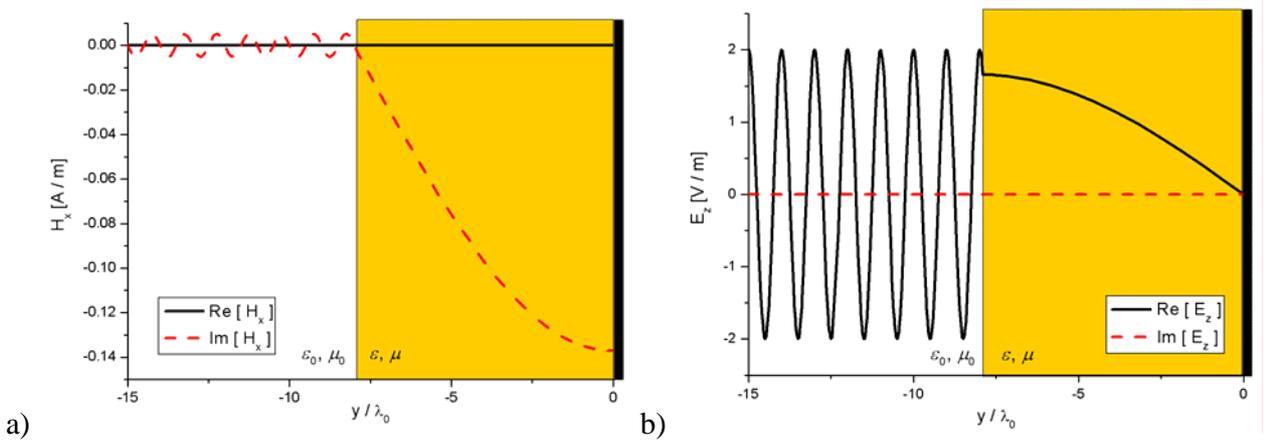

Figure 5 – Spatial variation of the tangential (a) magnetic and (b) electric field complex amplitudes versus y for an incident plane wave with $\mathbf{E}(x,y) = \hat{\mathbf{z}} e^{-jk_t x} e^{j\sqrt{k_0^2 - k_t^2} y}$ exciting the natural mode of the structure. The slab has the same parameters as in Figure 4.

The two plots clearly show how the material polariton of such a structure behaves in this case: the electric field inside the slab assumes values comparable with outside (i.e., twice the value of the incident field, due to the total reflection from the screen), whereas the magnetic field builds up on the screen, consistently with what was obtained in Figure 4 for plane waves close to broadside. If the plane wave had impinged at a different angle without noticeably exciting the polariton, we would have expected a much lower value for the electric field inside the slab (due to the low intrinsic impedance η of the chosen metamaterial) and a value of the magnetic field comparable with outside. Not surprisingly, such an excitation would not cause any enhancement in the transmission properties of the hole.

If we had shown a similar plot, but for a slab designed from formula (5) at the polariton resonance, the field distribution would have been dual with respect to this case: the electric field would have been enhanced in this case and not the magnetic field. The peak of the electric field, however, would have been at the centre of the cover slab ($\lambda/4$ apart from the screen) and therefore this resonant excitation would not be of any use in enhancing the hole transmission.

It is clear from this example that it is indeed true that an enhancement of transmission may be caused by the resonant excitations of surface plasmons [18] and leaky waves [16], but *the enhancement is not guaranteed by their presence*. The further condition, which should be explicitly mentioned, is that these natural modes should have their electromagnetic field concentrated as much as possible *on* the screen, in order to couple with the tiny aperture and to enhance its equivalent electric and magnetic dipoles. The excitation of this type of natural modes is crucial in obtaining a more significant and optimized enhancement of the wave transmission.

Similar results may be shown for the TM polarization, even if in this case we should also consider in general the contribution of the electric dipole in Eq. (1), corresponding to the possible presence of a normal electric field induced on the screen. For sake of brevity this case is not considered here.

### b) METAMATERIAL COVER IN THE EXIT FACE

Moving to the exit face of the screen, the considerations for an optimized design follow from those of the previous section applying the reciprocity theorem. The cover slab optimized for the entrance face, in fact, would support similar leaky waves also when placed at the exit side and these natural modes may have a double effect, namely, further enhancing the hole transmission and reshaping the beam towards the observer. The radiation expected on the exit side is in fact highly directive, due to the low damping factor of the leaky waves supported by the slab and, provided that the observer is placed where the radiation pattern shows its maximum, the received power may be significantly further enhanced. By reciprocity, the enhancement factor is again obtainable from Table 1 and if the observer is placed at the same angle from the normal to the screen as the impinging radiation at the entrance side, we expect to find the same enhancement factor, consistent with [16]-[17] (in those works both incident field and observer were placed at broadside). The total wave transmission in this case is increased by a factor $R_H^2$ and in terms of power by a factor $R_H^4$.

The radiation pattern for the equivalent electric and magnetic dipoles radiating from the region A at the exit side may be evaluated invoking again the reciprocity theorem, following e.g. [34]. Without loss of generality, we may suppose that the radiating magnetic dipole in A is directed along $\hat{\mathbf{x}}$, i.e., $\mathbf{m} = m\hat{\mathbf{x}}$, as it was for the impinging plane wave considered in the example of Figures 4 and 5. Applying this technique, its radiation pattern may be written in spherical coordinates as:

$$H_\theta(r, \theta, \phi) = \frac{k_0^2 m e^{-jk_0 r}}{4\pi r} H_0^{TE} \cos\phi$$
$$H_\phi(r, \theta, \phi) = -\frac{k_0^2 m e^{-jk_0 r}}{4\pi r} H_0^{TM} \sin\phi \quad , \qquad (7)$$

where $H_0^{TE}$ and $H_0^{TM}$ are the dimensionless complex amplitudes of the vectors given in (2), calculated for $c_0^{TE/TM} = \eta_0$ and $k_t = k_0 \sin\theta$ at $x = z = 0$ (where the dipole is supposed to be placed). Note that, unlike the standard spherical coordinate systems, here the coordinate system in (7) has its $\theta = 0$ axis along $\hat{\mathbf{y}}$ and its $\phi = 0$ axis along $\hat{\mathbf{x}}$.

Figure 6 shows the radiation patterns for the optimized setup, having covers as in Figure 4 both at the entrance and at the exit face. The pattern is compared to those obtained when no covers are present or when only the entrance or the exit covers are considered. At the exit face, the induced magnetic dipole excites both the TE and TM leaky waves supported by the exit cover; the TE pole has already been derived in the previous section and the TM one has $k_t^{TM} = (4.55 - 4.41j) \cdot 10^{-3} k_0$. The corresponding TM material polariton has $k_t^{TM} = 5.16 \cdot 10^{-3} k_0$. The total power enhancement achieved with both the covers, with respect to the hole without covers, is $R_H^4 = (\eta_0/\eta)^4 = 10^6 = 60\,\text{dB}$, as expected. The directivity when the exit cover is present is $D = 48.23\,\text{dB}$.

It is interesting to note that the thickness proposed in Figure 6 is not the one corresponding to the maximum directivity at broadside, since when $d_{out} = 0.991\pi/(2k)$, as found numerically, the directivity at broadside would be enhanced up to $48.78\,\text{dB}$. This is consistent with leaky wave antenna theory [28], since at broadside the maximum directivity is achieved when the leaky poles are slightly below cut-off. The same effect has been noticed by Oliner, Jackson and their co-workers [16]-[17] with their setup. However, the value adopted in Figure 6, which comes directly from formula (4), is the one that grants the maximum wave transmission for the given material. The fact that a higher directivity does not necessarily correspond to a higher enhancement of transmission for the broadside observer is due to the fact that the total power radiated by the magnetic dipole on the exit side is proportional to the magnetic field strength at the dipole location.

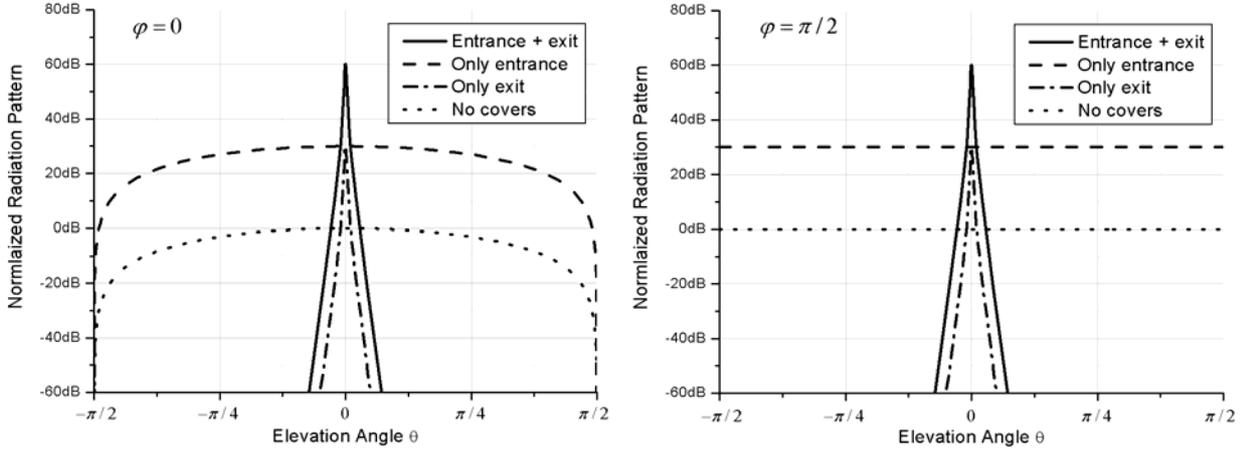

Figure 6: Radiation patterns in the E and H planes for covers with $\varepsilon = \varepsilon_0$, $\mu = 10^{-3}\mu_0$, $d_{in} = d_{out} = \pi/2k$, normalized to the peak of the pattern with no covers, compared to those cases with only the entrance cover, only the exit cover and the case with no covers.

The maximum enhancement of this quantity is granted by the choice $d = \pi/(2k)$, as shown in the previous section. The problem we are addressing, in fact, is not to enhance the efficiency of our exit antenna, for which a more directive beam is the best solution, but to extract the maximum power from the tiny hole and then radiating it as much as possible towards the observer. This task is obtained when condition (4) is satisfied at both sides, as also justified by reciprocity.

For a material with a higher $\varepsilon$, the possibility of increasing the angle $\theta^*$ of maximum directivity arises, as previously pointed out. As an example, Figure 7 shows the radiation patterns for cover slabs designed to receive an impinging TE plane wave from an angle $\theta^* = \pi/4$ and then radiating towards an observer placed again at $\theta^* = \pi/4$ on the other side of the screen. In this case the chosen parameters are $\varepsilon = 100\varepsilon_0$, $\mu = \mu_0/100$, $d_{in} = d_{out} = \pi/\left(2\sqrt{k^2 - k_0^2 \sin^2\theta^*}\right) = \pi/\left(\sqrt{2}\,k_0\right)$. The corresponding leaky poles have $k_t^{TE} \simeq k_t^{TM} = \left(1 - 6.36 \cdot 10^{-3} j\right) k_0 \sin\theta^*$ and the material polaritons have $k_t^{TE} = (1.004 - 0.06j) k_0 \sin\theta^*$ and $k_t^{TM} = 1.002 k_0 \sin\theta^*$. The corresponding power enhancement is $R^4 = \mu_0^4/\mu^4 = 10^8 = 80\,\text{dB}$ and the directivity at $\theta^* = \pi/4$ is $D = 19.77\,\text{dB}$, much lower than when it was at broadside, since the beam is now conical (it radiates along all the $\phi$ angles, due to the simultaneous excitation of the TE and TM leaky waves) and therefore by moving away from the broadside, the directivity decreases.

In this second example, it is evident how placing a slab on the exit side may not only have the effect of reshaping the beam in order to have it point towards our observer, as it was noticed to happen in the setup of [16]-[17], but it also causes *further enhancement of the total tunneled power*. As it is evident from the plot in Fig. 7, in fact, comparing the no exit-cover cases with those with the exit-cover, we find that the level of radiated power is enhanced at all angles (not only for $\theta = \theta^*$). This is due to the fact that the excited leaky wave at the exit face has been designed to have a strong magnetic field on the aperture and this couples very well with the equivalent magnetic dipoles and currents present there. The total radiated power is in fact proportional to the quantity $\mathbf{m} \cdot \mathbf{H}_0^*$, which is increased when natural modes that increase the amplitude of $\mathbf{H}_0$ are excited. In other words, in our setup we have shown that the exit side may play an active role in the enhancement effect, allowing also an increase of the total power radiated in the half-space region $y > 0$. This is due to the excitation of leaky-waves at the exit face with high magnetic field on the aperture, where the equivalent magnetic dipole representing the aperture is placed. Therefore, the total radiated power is enhanced by the presence of the exit cover, and moreover as a further effect the radiation is highly directive pointing towards the observer. For an optimized design, therefore, it becomes crucial to look for an exit cover that increases the value of the magnetic field on the slab exciting suitable natural modes, rather than a design that maximizes the directivity of the radiation towards the observer, as also previously noticed. This may also explain why our setup shows higher enhancement factors than those simulated in [16]-[17], where the corrugation responsible for the leaky-wave was made of metallic patches covering the screen, not specifically designed for the purpose of increasing the field on the metallic plane.

As an aside, it is also worth noting that when the cover material is isorefractive with the outside region, like in the case of Figure 7, the expression $R_H = \eta_0 / |\eta|$ is valid for every angle of incidence, as it may be also verified from the formulas in Table 1.

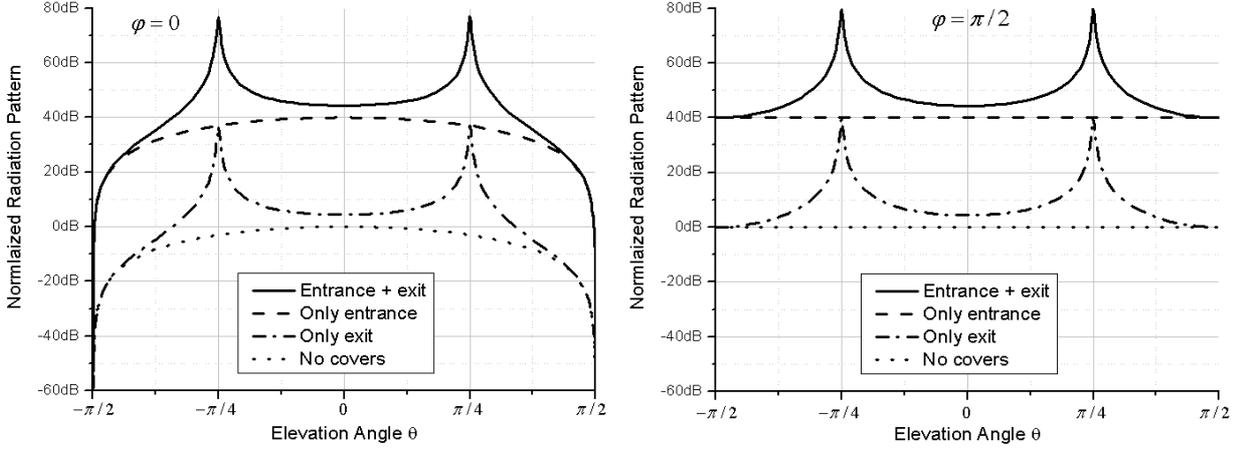

Figure 7: Radiation patterns in the E and H planes for covers with $\varepsilon = 100\varepsilon_0$, $\mu = \mu_0/100$, $d_{in} = d_{out} = \pi/(\sqrt{2}k_0)$, normalized to the peak of the pattern with no covers, compared to those cases with only the entrance cover, only the exit cover and the case with no covers.

## V. SENSITIVITY TO THE GEOMETRICAL AND ELECTROMAGNETIC PARAMETERS

In the previous sections, we showed that huge enhancements in the wave transmission through a sub-wavelength hole in a perfectly conducting screen can be obtained by using metamaterials with special parameters. Our simulations predict enhancements higher than the ones presented in the literature for different setups [14]-[19], [22]-[26], and the reasons for this effect have been pointed out. However, up to now we have considered the ideal situations in which the covers had exactly the required parameters, at the exact frequency of interest, no losses were considered, and we assumed infinite extension for the covers and infinite conductivity for the screen. In this section we discuss whether some of these requirements may be relaxed and how the performance is affected.

As a first parameter, we study the influence of material loss in our setup. When metamaterials with special parameters are employed, the losses may generally be not negligible and as a result may significantly affect the lossless results previously presented. Since the effects previously shown are closely related to the excitation of material polaritons on the screen surface, we expect that the presence of a reasonable amount of loss should not change the possibility of enhancing the wave transmission through the hole and the design formulas derived should still be valid. Of course, the total gain achieved may be lowered noticeably, but the phenomenon underlying the enhancement should still work, since lossy covers would still support material polaritons under similar conditions.

Figure 8 shows how the presence of loss may affect the enhancement phenomenon. In particular, the decrease of $R_H$ when the loss is increased in the setup of Figure 4 is reported in the left panel, while the change in the imaginary part of its TE leaky wave pole is reported in the right panel. The loss has been included by adding an imaginary part to the cover permeability, which is the most sensitive parameter, i.e., $\mu = \bar{\mu}(1-j\mu_i)$, where $\bar{\mu}$ is its real part, leaving unchanged the other parameters. As expected, the enhancement factor is sensitive to an increase in the losses, whereas the imaginary part of the complex poles is less affected. The sensitivity of the setup to the presence of loss depend in general on the ratio between real and imaginary part of the constitutive parameters. That is why a very low $\mu$, like the one required in the setup of Figure 4, may imply a very sensitive response of the enhancement performance to the presence of loss. In this sense, if the value of the imaginary part of $\mu$ is fixed, a relatively high value of $\bar{\mu}$ may become important to provide a good performance. Also, for the same reason the sensitivity of the permeability loss is expected to be much higher than the permittivity loss. It should be noted, however, that even an enhancement of $R_H = 10$ is a very good result (comparable with the best results obtained in [14]-[19], [22]-[26]), since in terms of power the total transmission enhancement is equal to its fourth power.

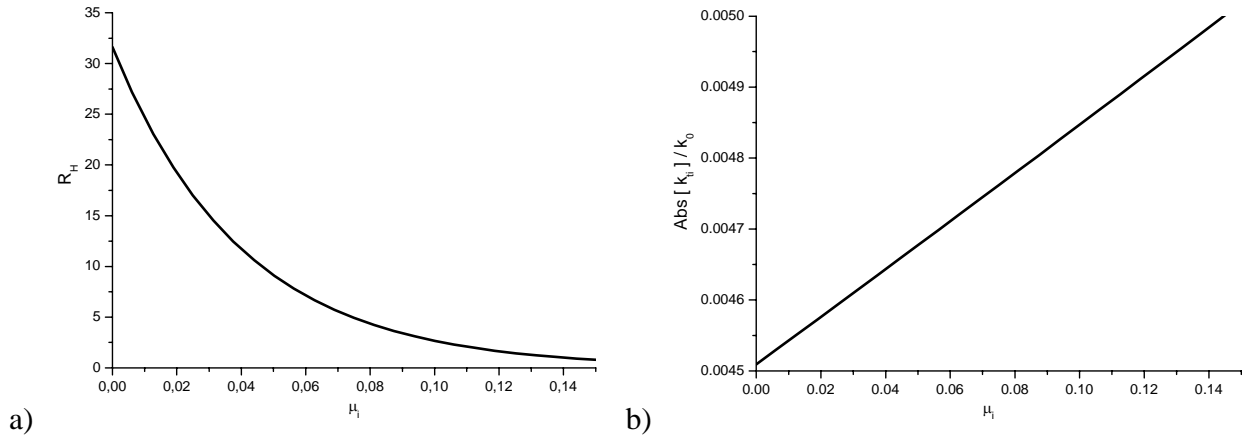

Figure 8: a) Variation of magnetic field enhancement and b) Variation of the magnitude of the imaginary part of the leaky pole, with respect to the imaginary part of the cover material permeability with real part $\bar{\mu} = 10^{-3}\mu_o$. The cover corresponds to the case of Figure 4.

A variation in frequency or a change in the thickness of the cover(s) have a similar effect on the performance of this setup, which can lead to change in the angle of maximum radiation, as it is evident from (4). It follows that such a setup may also be used as a spatial filter or a sort of prism, to radiate, at different angles, the different frequency components of a narrow-band signal impinging on the hole. This effect will not be further investigated in the present work, since it is similar to those verifiable in any leaky-wave setup.

The sensitivity to the variation of the constitutive parameters may also be considered here. In fact, the present technology may not always ensure a precise required bulk response for artificially engineered materials. In the case at hand, however, the effects on the enhancement are very clear from Table 1, and a variation in $\varepsilon$ and $\mu$ is not highly crucial in the wave transmission enhancement.

Also the finiteness of the cover in the transverse direction may be mentioned. If we move far enough away from the hole, the presence of the cover should not significantly affect the previous considerations and therefore its presence may not be required. This is due to the fact that, due to the imaginary part of the solutions for $k_t$, we expect a decay of the leaky waves propagating along the slab that are responsible for the enhancement effect. Of course, the higher the antenna directivity, the lower this decay is (since the leaky wave resembles more and more a plane wave distributed all over the transverse plane, and the imaginary part of $k_t$ is consequently lower and lower), and therefore the more sensitive the setup would be to a truncation of the cover at a given distance from the hole. However, it is a straightforward task for the designer to evaluate at which distance from the hole the cover is no longer essential, and considering losses this distance is further reduced. Investigation on the influence of the slab truncation in several leaky-wave antenna configurations have already been conducted by several researchers over the years (see e.g. [9], [27]) and the present case does not differ substantially.

Another parameter that may be considered in this analysis is the finiteness of the longitudinal thickness of the PEC screen. For values much smaller than the operating wavelength, which is usually the case at microwave frequencies, no sensible variation to the present analysis is expected. In general, however, the enhancement phenomenon may be further increased by the tunneling of energy through the little waveguide represented by the hole in this case. It should be noted, in fact, that since the hole has sub-wavelength transverse dimensions, the corresponding waveguide is below cut-off and therefore increasing the thickness of the screen (without covers) would cause the tunneling of energy to decrease exponentially. This effect has been theoretically predicted and experimentally verified in [36], [37]. However, the use of metamaterials with special parameters as covers at the two sides may induce an anomalous tunneling, which is similar in some ways to what was shown in [35]. It is interesting to note that also some researchers [38] have predicted and verified a similar tunneling in arrays of sub-wavelength holes in a conducting screen. In their work, they have derived a condition analogous to the one found in [35] for having this anomalous increase in the transmission through this otherwise opaque system, i.e., the condition of having two conjugate characteristic impedances at the two sides of the screen. The presence of material polaritons may give rise to this effect, which should play another important role in the transmission enhancement when the thickness of the screen is not negligible as compared to the wavelength of excitation. This case relates well with the considerations presented in the previous section on the active role played in the transmission enhancement by the exit side of our setup. Finally, a finite conductivity in the materials of the metallic screen should not change the conclusions here as much, provided that the screen thickness ensures that no coupling between the two sides through the metallic screen is present.

## VI. CONCLUSIONS

In this work, we have shown a detailed analysis and explanation of how metamaterial layers may be employed to enhance the wave transmission through a single sub-wavelength aperture in a perfectly conducting flat screen, when used as cover for the screen. First, we have compared this setup with the others proposed in the literature. Then, we have presented a complete and detailed theory explaining the enhancement phenomenon for our setup in terms of the leaky-wave theory and we have proposed optimal design formulas to improve the enhancement results. The theory presented here may also give further insights into the experimental results obtained by the other groups. Finally, we have studied the sensitivity of this design to losses and to variations in the geometrical and electromagnetic parameters of the setup.